\documentclass[preprintnumbers,aps]{revtex4}
\usepackage{graphicx}
\usepackage[latin1]{inputenc}
\usepackage{epsfig}
\usepackage{amsmath}
\usepackage{amsfonts}
\usepackage{float}
\usepackage{amssymb}%
\providecommand{\U}[1]{\protect\rule{.1in}{.1in}}
\def\unit{\leavevmode\hbox{\small1\kern-3.6pt\normalsize1}}
 \normalsize
\def\lsim{\raise0.3ex\hbox{$\;<$\kern-0.75em\raise-1.1ex\hbox{$\sim\;$}}}
\def\gsim{\raise0.3ex\hbox{$\;>$\kern-0.75em\raise-1.1ex\hbox{$\sim\;$}}}

\usepackage{graphics}
\def\bea{\begin{eqnarray}}
\def\eea{\end{eqnarray}}

\newcommand{\da}[1]{16\pi^2\;\frac{d #1}{dt}}
\newcommand{\be}{\begin{eqnarray}}
\newcommand{\ee}{\end{eqnarray}}

\begin{document}
\title{Radiative Symmetry Breaking in Supersymmetric $B-L$ Models with Inverse Seesaw}
\author{Shaaban Khalil}
\affiliation{Center for Fundamental Physics, Zewail City of Science and Technology,
Sheikh Zayed,12588, Giza, Egypt.}

\begin{abstract}
We study the radiative symmetry breaking of $B-L$ in supersymmetric models with  inverse seesaw mechanism. We show that for a wide region of parameter space the radiative corrections can drive the squared mass of the extra Higgs boson from positive initial values at the GUT scale to negative values at the TeV scale, leading to the spontaneous breaking of the $B-L$ symmetry. We also emphasize that in this class of models, unlike the supersymmetric $B-L$ models with  type I seesaw, the right-handed sneutrino cannot get a non-zero vacuum expectation value. Therefore, $B-L$ can be radiatively broken while $R$-parity remains an exact symmetry. 

\end{abstract}
\maketitle

\section{Introduction}
 The minimal $B-L$ extension of the Standard Model (SM), which is based on the gauge group $SU(3)_C \times SU(2)_L \times U(1)_Y \times U(1)_{B-L}$, provides a satisfactory explanation for the non-zero neutrino masses \cite{Khalil:2006yi,Basso:2008iv}.  In this class of models, SM singlet fermions are naturally introduced in order to cancel the  associated anomaly. These particles are accounted for right-handed neutrinos and hence a seesaw mechanism can be obtained. It was shown that light neutrino masses can be generated within $B-L$ extension of the SM  through either type-I seesaw  \cite{Khalil:2006yi} or inverse seesaw mechanism \cite{Khalil:2010iu}. In type-I seesaw mechanism right-handed neutrinos acquire Majorana masses at the $B-L$ symmetry breaking scale, therefore the neutrino's Yukawa coupling must be $\lsim {\cal O}(10^{-6})$, while in inverse seesaw these Majorana masses are not allowed by the $B-L$ gauge symmetry and another pair of SM gauge singlet fermions with tiny masses $\sim$ keV must be introduced. One of these two singlets fermions couples to right handed neutrino and is involved in generating the light neutrino masses. 
 
Furthermore, it was shown that in a SUSY context, the $B-L$ and SUSY scales can be correlated through the mechanism
of radiative breaking of $B-L$ symmetry, similarly to the radiative electroweak symmetry breaking in MSSM \cite{Khalil:2007dr}. 
In particular, it was proven that the radiative corrections in  $B-L$ extension of the MSSM (BLSSM) with type I seesaw may drive the squared
mass of extra Higgs boson from positive initial values at the GUT scale to negative values at the TeV scale, leading to spontaneous breaking of 
$B-L$. Thus, the energy scale of $B-L$ breaking is naturally related to the SUSY breaking scale. However, it was pointed out \cite{FileviezPerez:2010ek} that the breaking of $B-L$ in this model depends crucially on the large value of the right-handed neutrino Yukawa coupling and it is possible to break the $B-L$ through the Vacuum Expectation Value (VEV) of the right-handed sneutrino. In this case $R$-parity is also spontaneously broken and the resulting model will have quite involved phenomenology. 

In this paper we analyze the radiative $B-L$ symmetry breaking in BLSSM with Inverse Seesaw (BLSSM-IS).  We show that the breaking of $B-L$ occurs for a wider region of parameter space through the VEV of the Higgs singlet. We consider the Renormalisation Group Equations (RGEs) to show explicitly that for wide range of parameters the squared mass of the Higgs singlet can be negative at TeV scale while the squared mass of the right-handed sneutrino remains positive. Therefore, the $B-L$ symmetry is spontaneously broken by the VEV of this singlet and $R$-parity remains exact. In addition, using the program of Vevacious \cite{Camargo-Molina:2013qva}, we analyze the vacuum stability in both BLSSM-IS and  BLSSM-type I. We show that, unlike the BLSSM-type I, in BLSSM-IS the VEV of right-handed sneutrino is always close to zero and much less than the VEV of the singlet scalar that breaks the $B-L$ and keeps $R$-party conserved.  

The plan of  the paper is as follows. In the next section, we analyze the RGE running and the radiative $B-L$ symmetry breaking in BLSSM with inverse seesaw and compare it with the results of the BLSSM with type I seesaw.  In Section 3 we investigate the vacuum stability in the BLSSM-IS and also in BLSSM-type I. We conclude in Section 4.

\section{RGE running and $B-L$ symmetry breaking}

TeV scale BLSSM-IS is based on the gauge group $SU(3)_C\times SU(2)_L\times U(1)_Y\times
U(1)_{B-L}$, where the $U(1)_{B-L}$ is spontaneously broken by 
chiral singlet superfields $\chi_{1,2}$ with $B-L$ charge $=\pm 1$
As in conventional $B-L$
model, a gauge boson $Z_{BL}$ and three chiral singlet
sueperfields $\nu_{R_i}$ with $B-L$ charge $=-1$ are introduced for
the consistency of the model. Finally, three chiral singlet
superfields $S_1$ with $B-L$ charge $=+2$ and three chiral singlet
superfields $S_2$ with $B-L$ charge $=-2$ are considered to
implement the inverse seesaw
mechanism \cite{Khalil:2010iu}. The superpotential of the leptonic sector of this model is given by%
\bea%
{\cal W}= Y_{e}\, E^c L H_1 + Y_{\nu}\, \nu_R^c LH_2  + Y_{S}\, \nu_R^c \chi_1
S_{2}+ \mu\, H_1H_2 + \mu'\, \chi_1\chi_2.
\label{sp}
\eea
Note that the chiral singlet superfields $\chi_2$ and $\nu_R^c$ have the same $B-L$ charge. Therefore, one may impose a discrete symmetry in order to distinguish them and to prohibit other terms beyond those given in Eq. (\ref{sp}).
In this case, the relevant soft SUSY breaking terms, assuming the usual
universality assumptions, are as follows
 \bea - {\cal L}_{soft} &=&
\sum_{\phi} {\widetilde{m}}_{\phi}^{2} \vert \phi \vert^2 +
Y_{\nu}^{A}{\widetilde{\nu}_R}^{c} {\widetilde{L}}H_{2} +
Y_{e}^{A}{\widetilde{E}}^{c}{\widetilde{L}}H_{1}
+Y_{S}^{A}{\widetilde{\nu}_R}^{c}{\widetilde
S_{2}} \chi_1  +B \mu H_1 H_2 + B\mu' \chi_1 \chi_2
 \nonumber\\&+& \frac{1}{2} M_1 {\widetilde{B}}{\widetilde{B}}+
 \frac{1}{2}M_2{\widetilde{W}}^a {\widetilde{W}}^a
+ \frac{1}{2}M_3 {\widetilde{g}}^a {\widetilde{g}}^a + \frac{1}{2}
M_{BL} {\widetilde{Z}_{BL}}{\widetilde{Z}_{BL}}+ h.c ,
\label{soft}
\eea%
where the sum in the first term runs over $\phi=H_1, H_2, \chi_1,
\chi_2, \tilde L, \tilde E^c,\tilde{\nu}_R^c,\tilde S_1, \tilde S_2$
and $Y_L^A\equiv Y_L A_L$ ($L=e,\nu,S$) is the trilinear scalar interaction coupling associated with lepton
Yukawa coupling.  In order to prohibit a possible large mass term
$M S_1 S_2$ in the above, we assume that the particles, $\nu^c_{R_i}$,
$\chi_{1,2}$, and $S_2$ are even under matter parity, while $S_1$
is an odd particle. The $B-L$ symmetry can be radiatively broken by
the non-vanishing vacuume expectation values (VEVs) $\langle
\chi_1 \rangle = v'_1$ and $\langle \chi_2 \rangle = v'_2$
\cite{Khalil:2007dr}. The tree level
potential $V(\chi_1,\chi_2)$ is given by %
\begin{eqnarray}
V( \chi_1, \chi_2 ) = \mu_1^2 |\chi_1|^2 + \mu_2^2 |\chi_2|^2 - \mu_3^2 ( \chi_1 \chi_2 + h.c. ) + \frac{1}{2} g_{BL}^2 \left( |\chi_2|^2 - |\chi_1|^2 \right)^2,
\end{eqnarray}
where $\mu^2_{1,2} = m^2_{\chi_{1,2}}+ \vert \mu' \vert^2$ and $\mu_3^2=- B' \mu'$.
The stablitity condition of $V(\chi_1, \chi_2)$ is given by
\begin{equation}
2 \mu_3^2 < \mu_1^2 + \mu_2^2.
\label{4}
\end{equation}
A non-zero minimum may be obtained if there is a negative squared mass eigenvalue in the $B-L$ Higgs mass matrix, {\it i.e.}, if 
\begin{equation}
\label{6}
\mu_1^2\; \mu_2^2 < \mu_3^4.
\end{equation}
this condition is not satisfied at the GUT scale with universal soft breaking terms. However, as we will show, similar to the MSSM scalar Higgs masses, the running from a large scale down to TeV scale, $\mu^2_1$ and $\mu_2^2$ will have different renormalization scales so that the minimization
condition is  eventually satisfied, and hence, the $B-L$ symmetry is spontaneously broken.  
The minimization conditions, $\frac{\partial V}{\partial\chi_i}=0,\ i=1,2$, lead to the following equations:
\begin{eqnarray}
\vert \mu'\vert ^2 &=& \frac{m_{\chi_2}^2 - m_{\chi_1}^2 \tan \beta'}{\tan \beta' -1} - M_{Z'}^2/2 ,\label{tadapole1}\\
\sin 2 \beta' &=& \frac{-2 B' \mu' }{m_{\chi_1}^2 + m_{\chi_2}^2 + 2 \vert \mu \vert^2},
\label{tadapole2}
\end{eqnarray}
where $\tan \beta' =v_1/v_2$ and $M^2_{Z_{BL}}=4 g^2_{BL} (v_1^2+v_2^2)$.
These two equations are similar to the electroweak symmetry breaking conditions in MSSM which are used to determine the value of $\mu$ and $B$ parameters at the electroweak scale. It is worth noting that in MSSM, where $\tan \beta > 1$, one cannot satisfy the condition of non-vanishing VEVs and $\vert \mu\vert^2 > 0 $ unless the running  from GUT to weak scale reduces $m_{H_u}^2$ to negative values, thanks  to the large Yukawa coupling of $H_u$ with top quark. The situation with $B-L$ symmetry breaking could be different. The conditions $\vert \mu' \vert^2 >0$ and $\mu_1^2 \mu_2^2 < \mu_3^4$ can be simultaneously satisfied with positive $m_{\chi_{1,2}}^2$, if $\tan \beta' \sim 1$. Before elaborating this point, let us  consider the running of the scalar masses $m_{\chi_{1,2}}^2$ and also the right-handed sneutrino squared masses, $m_{\tilde{\nu}_R}^2$, via the $B-L$ Renormalization Group Equations (RGEs) in both type I and inverse seesaw mechanisms.

In type I seesaw, these RGEs are given by 
\bea
\da{m_{\chi_1}^2} & = & -12 g_{BL}^2 {M}_{BL}^2 + 4 m_{\chi_1}^2 {\rm Tr}(Y_{\nu_R} Y_{\nu_R}^*) + 4 {\rm Tr}(Y^{A^*}_{\nu_R} Y^A_{\nu_R}) + 8 {\rm Tr}(m^2_{\tilde{\nu}_R}  Y_{\nu_R} Y^*_{\nu_R}),\\
\da{m_{\tilde{\nu}_R}^2} & = & -3 g_{BL}^2 {M}_{BL}^2 +8 m_{\chi_1}^2  Y_{\nu_R} Y_{\nu_R}^* + 8 Y^{A^*}_{\nu_R} Y^A_{\nu_R} + 4 m^2_{\tilde{\nu}_R} Y_{\nu_R} Y_{\nu_R}^* 
+ 8 Y_{\nu_R} m^2_{\tilde{\nu}_R} Y_{\nu_R}^* + 4 Y_{\nu_R} Y_{\nu_R}^* m^2_{\tilde{\nu}_R} ,
\eea
where $Y_{\nu_R}$ is the Yukawa coupling of right-handed neutrino term in type I superpotential: $Y_{\nu_R} \nu_R^c \chi_1 \nu_R$ and the trilinear coupling $T_{\nu_R}$ is defined as usual as $Y^A_{\nu_R}= Y_{\nu_R} A_{\nu_R}$. Thus for $Y_{\nu_R} = Y_{\nu_R} ~{\rm diag}\{0,0,1\}$ one finds
 \bea
\da{m_{\chi_1}^2} & = & -12 g_{BL}^2 {M}_{BL}^2 + 4 Y^2_{\nu_R} \left(m_{\chi_1}^2  + A_{\nu_R}^2 + 2 m^2_{\tilde{\nu}_R}\right ),\\
\da{m_{\tilde{\nu}_R}^2} & = & -3 g_{BL}^2 {M}_{BL}^2 +8  Y^2_{\nu_R} \left(m_{\chi_1}^2 + A_{\nu_R}^2 + 2 m^2_{\tilde{\nu}_R}\right).
\eea
The last term proportional to $Y_{\nu_R}$ in these equations derives the mass squared negative at TeV scale. Therefore, is clear that $m_{\tilde{\nu}_R}^2$ can be negative before $m_{\chi_1}^2$ (due to the large coefficient of $Y_{\nu_R}$ in the RGE of $m_{\tilde{\nu}_R}^2$). In this case of hierarchal $Y_{\nu_R}$, both $B-L$ and $R$-parity will be spontaneously broken \cite{FileviezPerez:2010ek}. However, in case of $Y_{\nu_R} = Y_{\nu_R} ~{\rm diag}\{1,1,1\}$, the equations take the form  
 \bea
\da{m_{\chi_1}^2} & = & -12 g_{BL}^2 {M}_{BL}^2 + 12 Y^2_{\nu_R} \left(m_{\chi_1}^2  + A_{\nu_R}^2 + 2 m^2_{\tilde{\nu}_R}\right ),\\
\da{m_{\tilde{\nu}_R}^2} & = & -3 g_{BL}^2 {M}_{BL}^2 +8  Y^2_{\nu_R} \left(m_{\chi_1}^2 + A_{\nu_R}^2 + 2 m^2_{\tilde{\nu}_R}\right).
\eea
Therefore it is expected that $m_{\chi_1}^2$ becomes negative at TeV scale while $m_{\tilde{\nu}_R}^2$ remains positive, so $B-L$ symmetry is spontaneously broken and $R$-parity remains exact \cite{Khalil:2007dr}. 

\begin{figure*}[t!]
\begin{center}
\includegraphics[width=8.25cm, height=6.5cm]{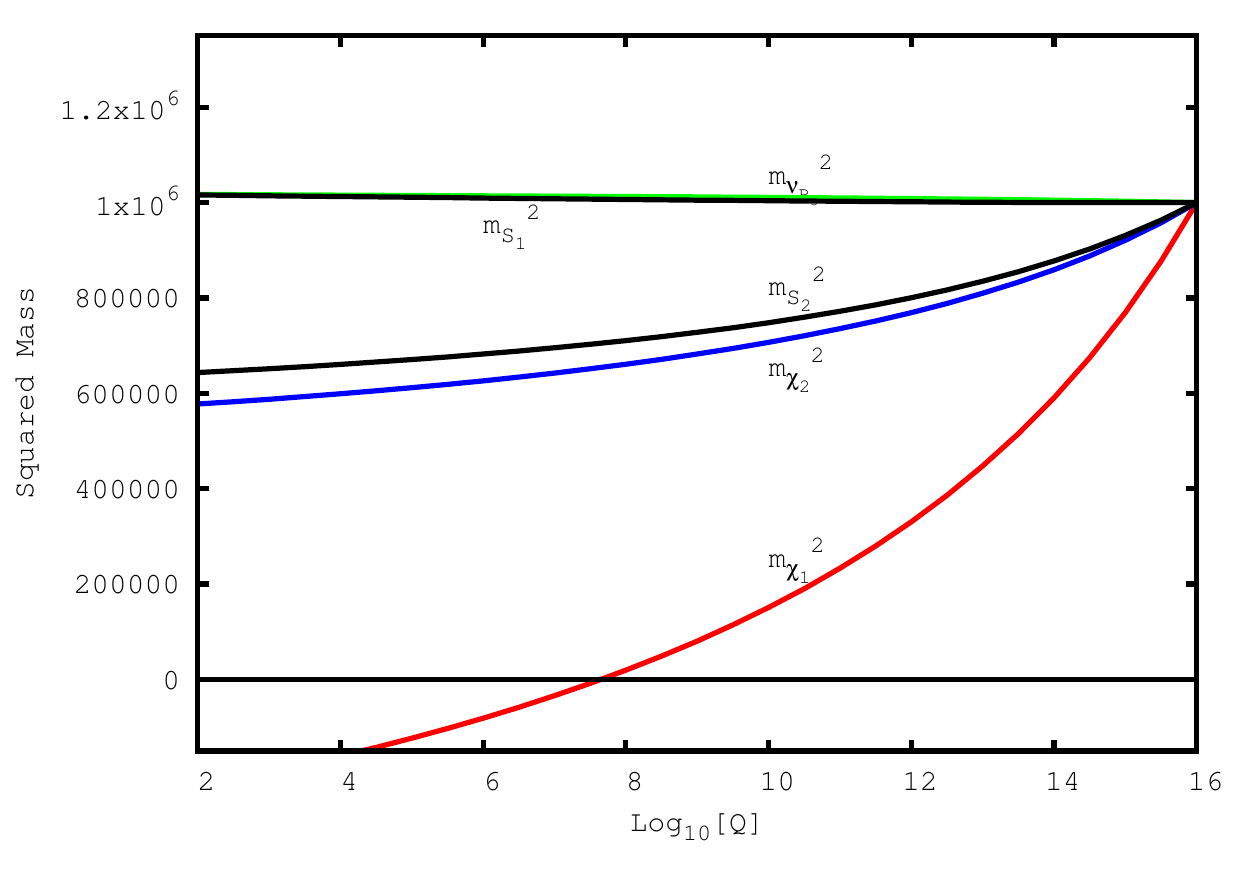}~~ \includegraphics[width=8.25cm, height=6.5cm]{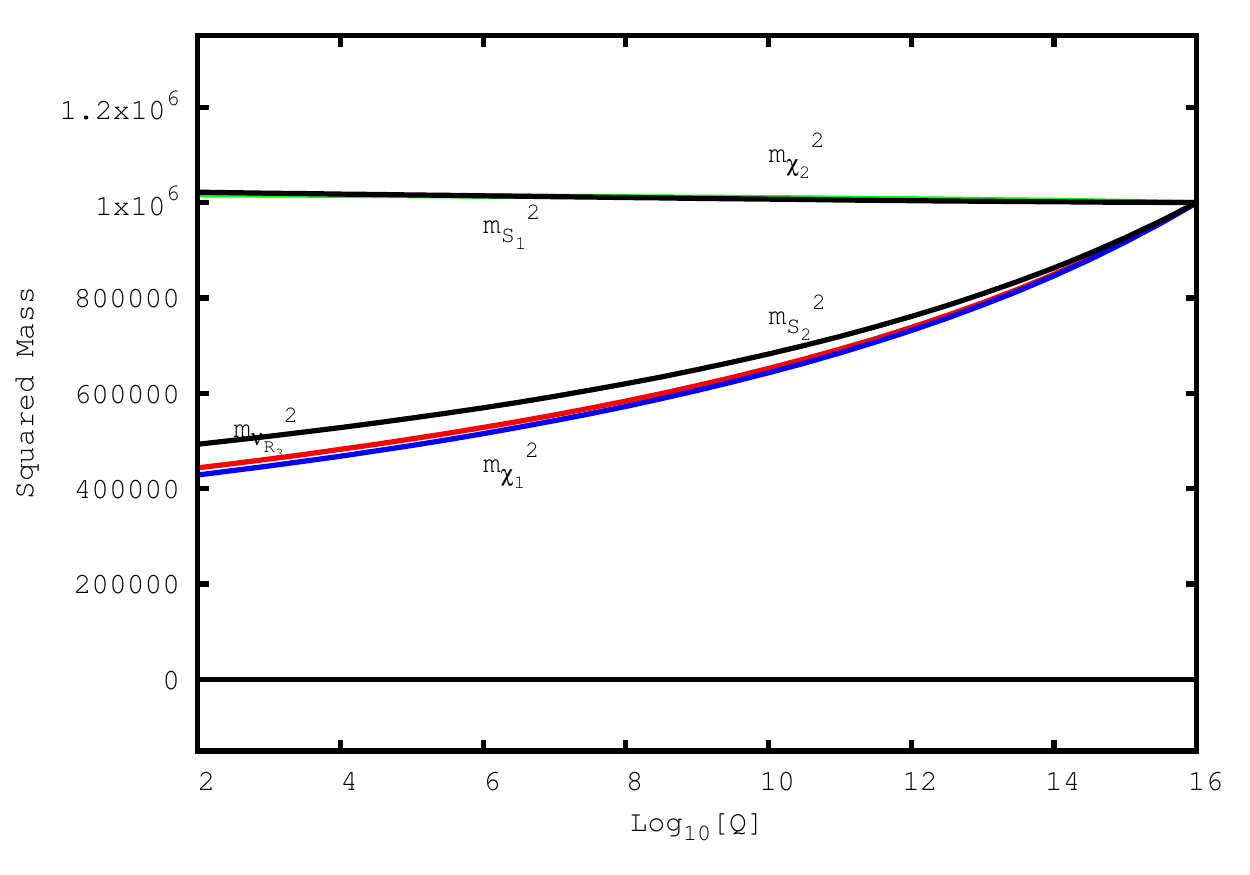}\\
\includegraphics[width=8.25cm, height=6.5cm]{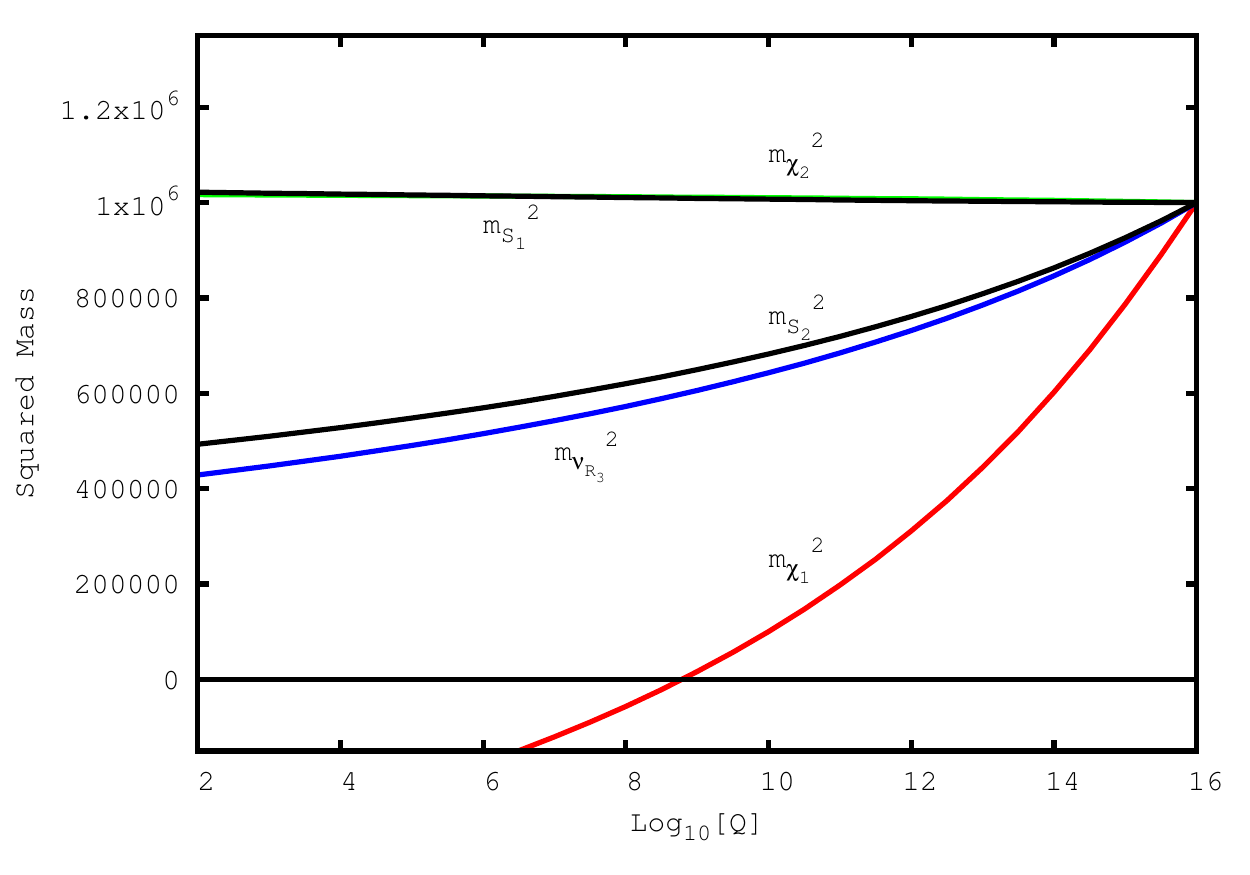}~~ \includegraphics[width=8.25cm, height=6.5cm]{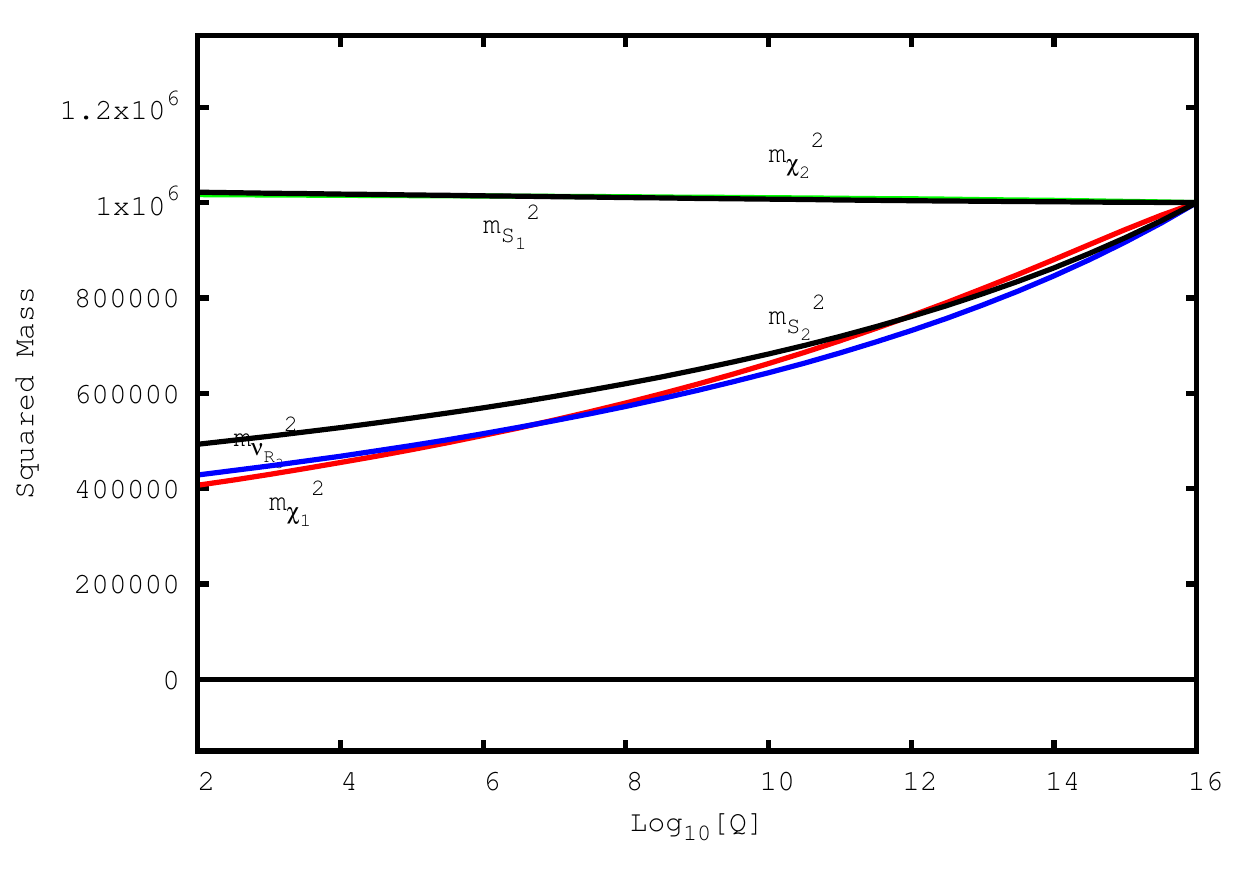}
\caption{The evolution of the $B-L$ scalar masses: $m_{\chi_1}^2$ (red),  $m_{\chi_2}^2$ (black ), and $m_{\tilde{\nu}_R}^2$ (blue) in BLSSM with inverse seesaw from GUT to TeV scale for $m_0 =m_{1/2}=A_0= 1$ TeV,  $Y_{s}\sim {\rm diag}\{1,1,1\} $ and $Y_{\nu}\sim {\cal O}(0.1) \times {\rm diag}\{1,1,1\}$ (left) and $Y_{s}\sim {\rm diag}\{0,0,1\} $ and $Y_{\nu}\sim {\cal O}(0.1) \times {\rm diag}\{0,0,1\}$ (right), and $g_{BL} \sim 0.1$ (up) and $g_{BL} \sim 0.5$ (down).}
\label{RGE-BLSSMIS}
\end{center}
\end{figure*}

In inverse seesaw, the relevant RGEs are given by 
\bea
\da{m_{\chi_1}^2} & = & -12 g_{BL}^2 {M}_{BL}^2 + 2 m_{\chi_1}^2 {\rm Tr}(Y_{s} Y_{s}^\dag) + 2 {\rm Tr}(Y^{A^*}_{s} Y^{A^T}_{s}) + 2 {\rm Tr}(m^2_{\tilde{s}_2}  Y^\dag_{s} Y_{s}) + 2 {\rm Tr}(m^2_{\tilde{\nu}_R} Y_\nu Y^\dag_\nu),\\
\da{m_{\chi_2}^2} & = & -12 g_{BL}^2 {M}_{BL}^2 ,\\
\da{m_{\tilde{\nu}_R}^2} & = & -3 g_{BL}^2 {M}_{BL}^2 +2 m_{\chi_1}^2  Y_{s} Y_{s}^\dag + 2 Y^A_{s} Y^{A^\dag}_{s} + 4 Y^A_{\nu} Y^{A^\dag}_{\nu}+  m^2_{\tilde{\nu}_R} Y_{s} Y_{s}^\dag + 2  m^2_{\tilde{\nu}_R} Y_{\nu}Y_{\nu}^\dag \nonumber\\
&+& 2 Y_{s} m^2_{\tilde{s}_2} Y_{s}^\dag + Y_s  Y_{s}^\dag m^2_{\tilde{\nu}_R} + 2  Y_{\nu}Y_{\nu}^\dag m^2_{\tilde{\nu}_R} ,\\
\da{m_{\tilde{s}_2}^2} & = & -3 g_{BL}^2 {M}_{BL}^2 +2 m_{\chi_1}^2  Y_{s}^\dag Y_{s} + m^2_{\tilde{s}_2} Y_s^\dag Y_s + 2 Y^{A^\dag}_{s} Y^{A}_{s} + 2 Y^\dag_{s} m_{\tilde{\nu}_R}^2 Y_{s}+  Y_{s}^\dag Y_{s} m^2_{\tilde{s}_2}.
\eea
Thus for hierarchical Yukawas: $Y_{s} = Y_{s} ~{\rm diag}\{0,0,1\}$ and $Y_{\nu} = Y_{\nu} ~{\rm diag}\{0,0,1\}$, one gets
\bea
\da{m_{\chi_1}^2} & = & -12 g_{BL}^2 {M}_{BL}^2 + 2Y_{s}^2 \left( m_{\chi_1}^2 + A_s^2 +  m_{\tilde{s}_{2}}^2 \right) + 2 m_{\tilde{\nu}_R}^2 Y_\nu^2,\\
\da{m_{\tilde{\nu}_R}^2} & = & -3 g_{BL}^2 {M}_{BL}^2 + 2 Y_{s}^2 \left( m_{\chi_1}^2 +  A_s^2 + m^2_{\tilde{s}_2}+ m_{\tilde{\nu}_R}^2\right) + 
4 Y_\nu^2 (m_{\tilde{\nu}_R}^2 +  A_\nu^2),\\
\da{m_{\tilde{s}}^2} & = & -3 g_{BL}^2 {M}_{BL}^2 + 2 Y_{s}^2 \left( m_{\chi_1}^2 +   m_{\tilde{s}_{2}}^2 + A_{s}^2+ m_{\tilde{\nu}_R}^2 \right) .
\eea
From these equations, one can see that in inverse seesaw scenario the evolution of $m_{\chi_1}^2$ and $m_{\tilde{\nu}_R}^2$ depends on the relative strength of $Y_s$ and $Y_\nu$. In the case of hierarchal Yukawa,  $m_{\chi_1}^2$ can be of order or slightly smaller than $m_{\tilde{\nu}_R}^2$ if $Y_\nu \ll Y_s$. From the RGEs of $Y_s$ and $Y_\nu$ one can notice that $Y_\nu$ must be $\lsim 0.5$, to avoid a possible Landau pole at high scale, while $Y_s$ can be of order one. In these conditions, the masses $m_{\chi_1}^2$ and $m_{\tilde{\nu}_R}^2$ are of the same order and positive (as shown explicitly in Fig. \ref{RGE-BLSSMIS}). So that $R$-parity remains exact symmetry and $B-L$ could be broken if symmetry breaking conditions in Eq.~\ref{tadapole2} are satisfied.  As intimated, these conditions do not require negative mass squared scalar masses and with $\tan \beta' \simeq {\cal O}(1)$, symmetry can be broken with $m_{\chi_1}^2 >0$. So even if $ m_{\tilde{\nu}_R}^2 \leq m_{\chi_1}^2$ and both are positive, $B-L$ symmetry only can be broken. In case of degenerate Yukawa, {\it i.e.},  $Y_{s} = Y_{s} ~{\rm diag}\{1,1,1\}$ and $Y_{\nu} \sim {\cal O}(0.5) \times {\rm diag}\{1,1,1\}$, the splitting between $m_{\chi_1}^2$ and $m_{\tilde{\nu}_R}^2 ~\&~ m_{\tilde{s}}^2$ becomes larger, so that $m_{\chi_1}^2$ can be negative while $m_{\tilde{\nu}_R}^2$ and $m_{\tilde{s}_2}^2$ are positive. 

%

In Fig. \ref{RGE-BLSSMIS} we display the scale evolution of the Higgs masses $m_{\chi_{1,2}}^2$ and also the scalar masses $m_{\tilde{\nu}_{R_3}}^2$ and $m_{\tilde{S}_{1,2}}^2$ based on  the numerical solution of  complete RGEs derived by using SARAH \cite{Staub:2013tta}, for  $m_0 = M_{1/2}= A_0=1$ TeV and $Y_{\nu} \sim 1$ is assumed. As can be seen from this figure, $m^2_{\chi_1}$ drops rapidly to the negative region, while $m^2_{\chi_2}$ and other scalar masses remain positive at TeV scale.  Analogously to the radiative electroweak symmetry breaking in MSSM, this mechanism works with large Yukawa coupling.  It is worth mentioning that unlike the type I seesaw BLSSM, here the scalar mass $m_{\tilde{\nu}_{R_3}}^2$ remains positive at the low scale independently of the initial values. Hence the $B-L$ breaking via a non-vanishing VEV for right-handed sneutrinos $\tilde{\nu}_{R_3}$, does not occur in the present framework and $R$-parity remains exact.  As intimated, if right-handed sneutrino acquires a non-vanishing VEV, then both $B-L$ and $R$-parity would simultaneously be broken. In this case, the model leads to a quite different and involved phenomenology at the low scale \cite{FileviezPerez:2010ek,Kikuchi:2008xu,Fonseca:2011vn}. 

\begin{figure*}[t!]
\begin{center}
\includegraphics[width=8.25cm, height=6.5cm]{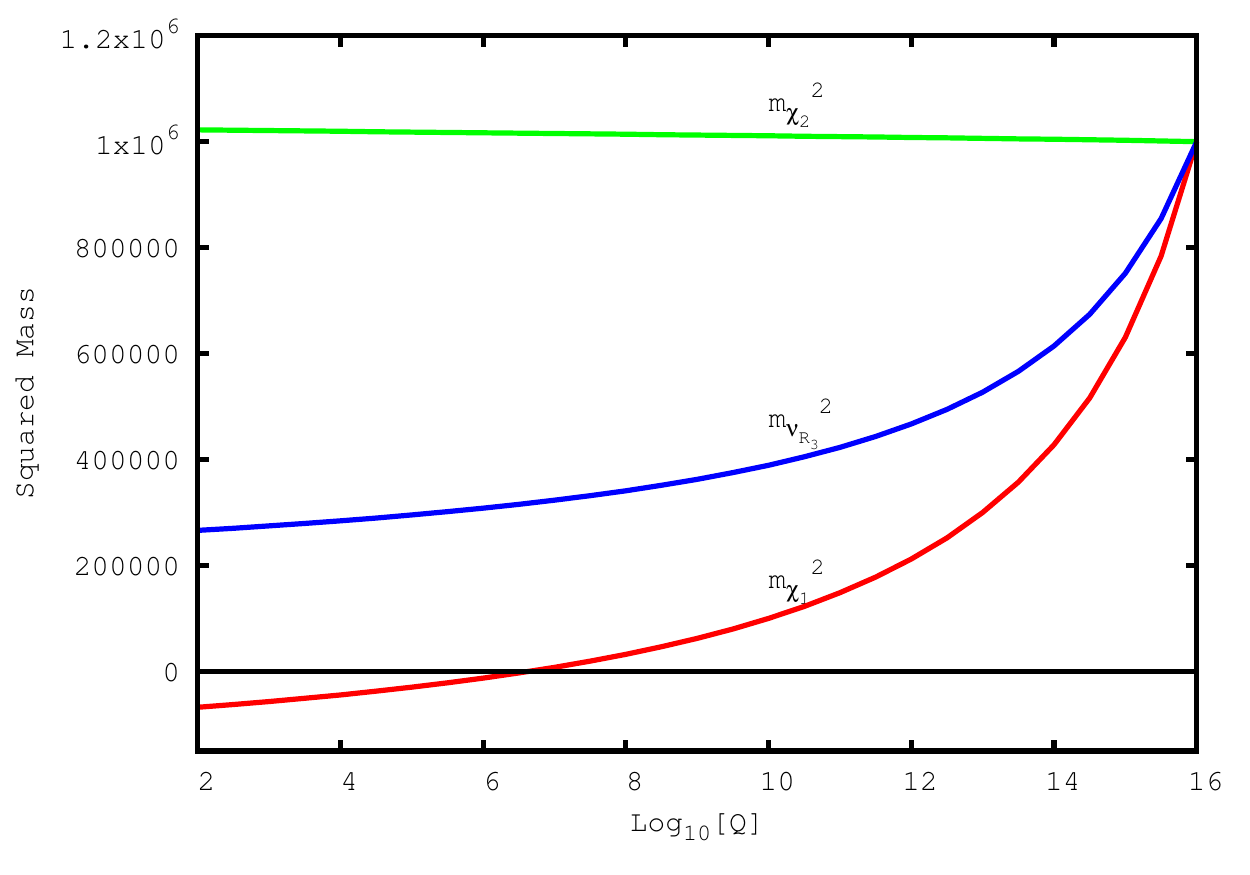}~~ \includegraphics[width=8.25cm, height=6.5cm]{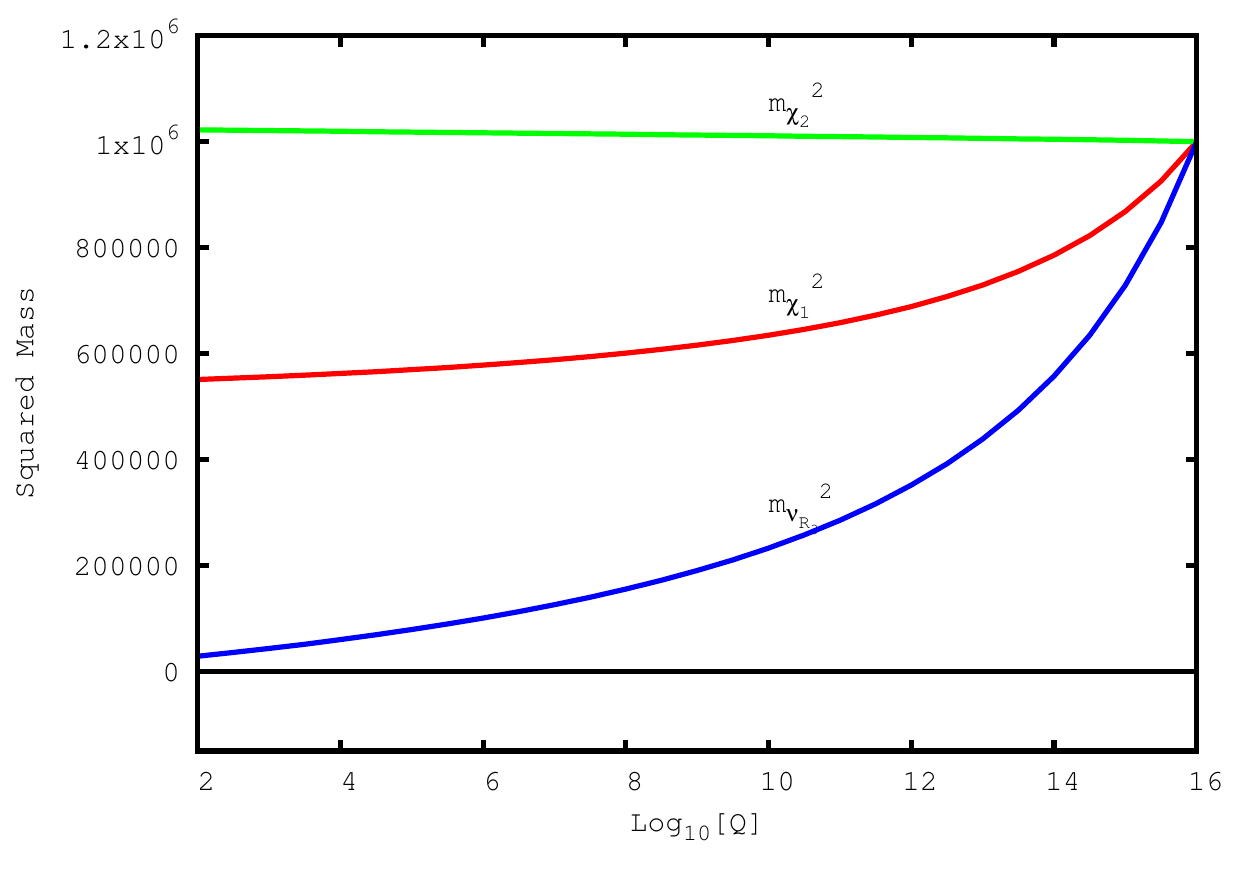}\\
\includegraphics[width=8.25cm, height=6.5cm]{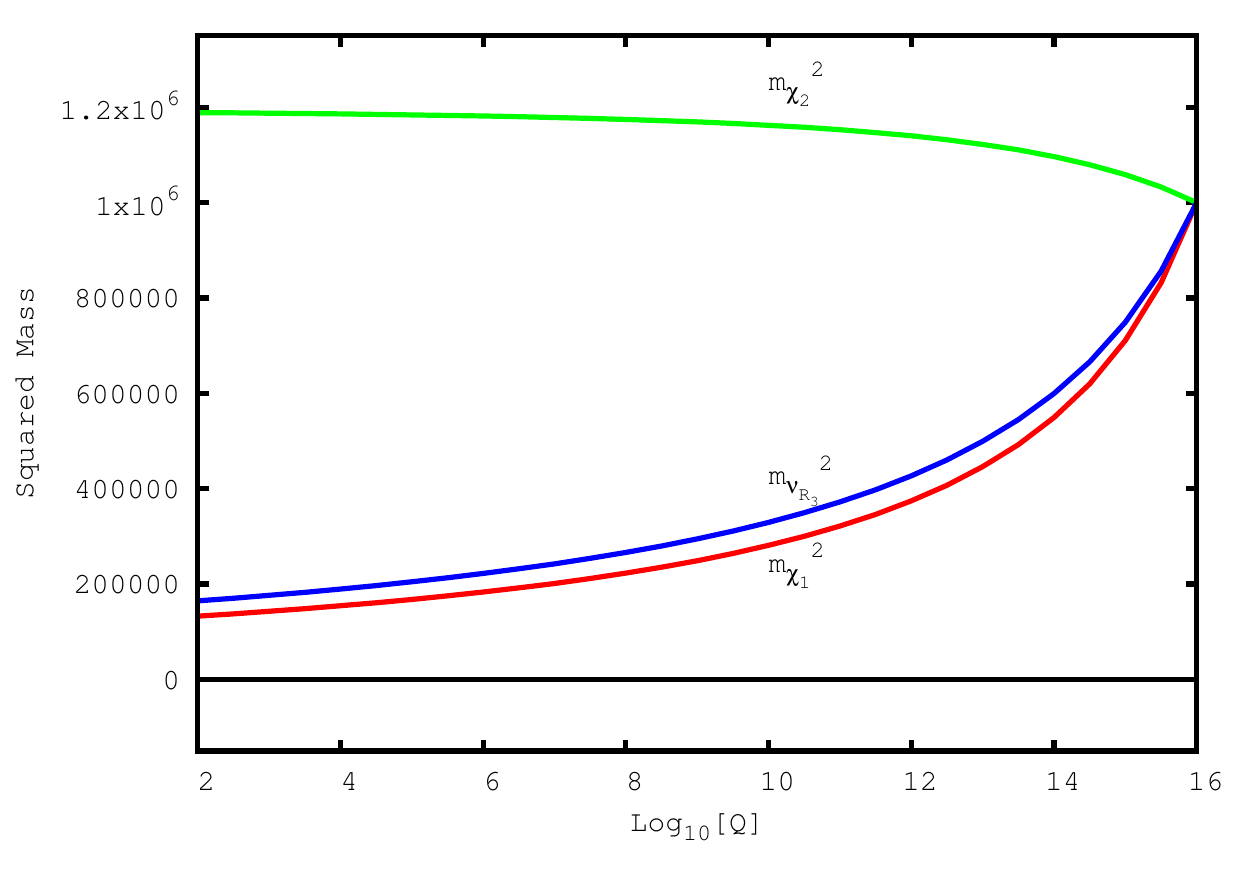}~~ \includegraphics[width=8.25cm, height=6.5cm]{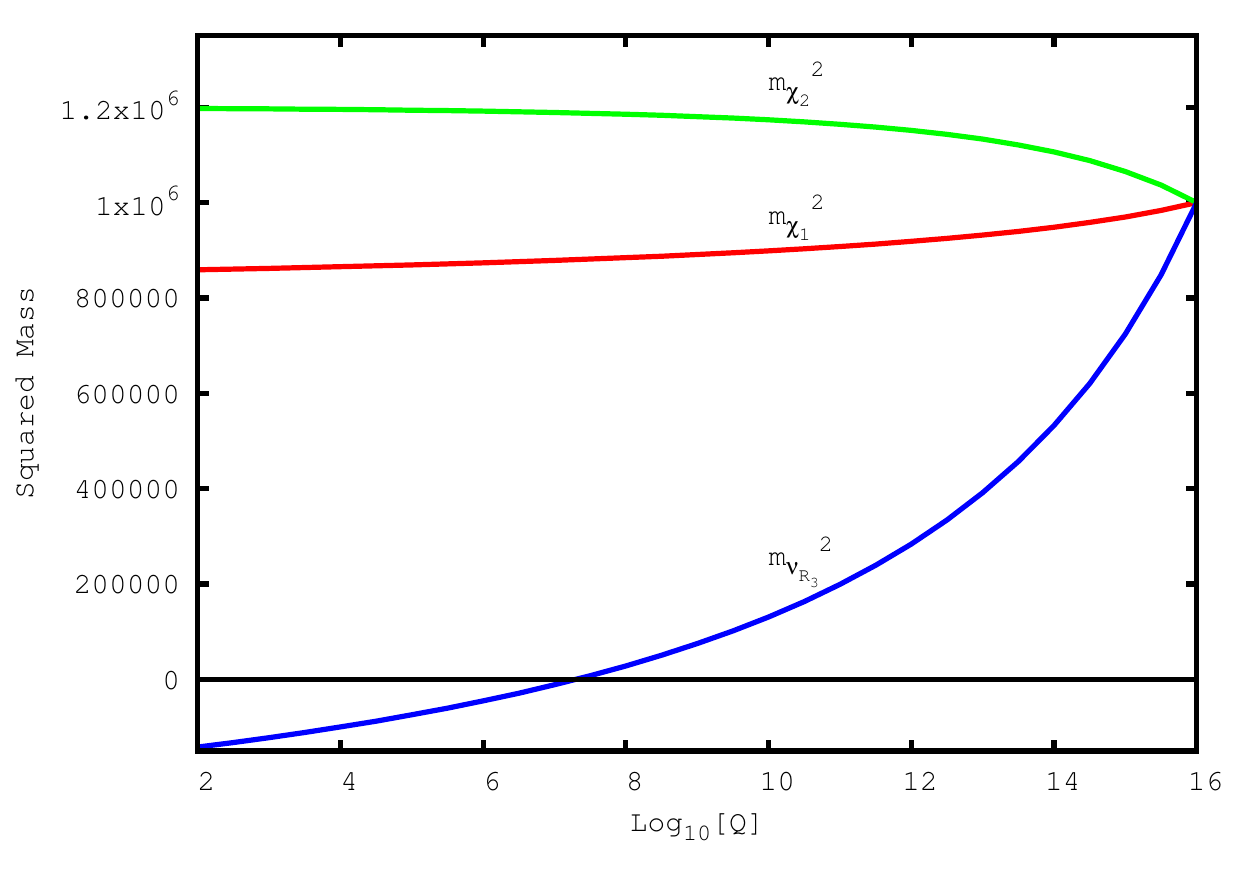}
\caption{The evolution of the $B-L$ scalar masses: $m_{\chi_1}^2$ (red),  $m_{\chi_2}^2$ (black ), and $m_{\tilde{\nu}_R}^2$ (blue) in BLSSM with type I seesaw from GUT to TeV scale for $m_0 =m_{1/2}=A_0= 1$ TeV,  $Y_{\nu}\sim {\rm diag}\{1,1,1\}$ (left) and $Y_{\nu}\sim {\rm diag}\{0,0,1\}$ (right), and $g_{BL} \sim 0.1$ (up) and $g_{BL} \sim 0.5$ (down).}
\label{RGE-BLSSM}
\end{center}
\end{figure*}
%

\section{$B-L$ Vacuum stability}

In this section we analyze, by using Vevacious \cite{Camargo-Molina:2013qva}, the vacuum stability of the BLSSM-IS. We perform a wide scan over all relevant parameters and calculate the scalar potential through SARAH \cite{Staub:2013tta} and SPheno \cite{Porod:2003um}. The stability results are classified into two categories: $(i)$ Non-vanishing VEVs of $\chi_i$: $\langle \chi_i \rangle \neq 0 ~ (\sim 10^3)$ GeV and $\langle \tilde{\nu}_{R_3} \rangle = 0$, where $B-L$ is spontaneously broken and $R$-parity is conserved. $(ii)$ $\langle \chi_i \rangle =0$ and $\langle \tilde{\nu}_{R_3} \rangle \neq 0~ (\sim 10^3)$ GeV, so that both $B-L$ and $R$-parity are spontaneously broken.

\begin{figure*}[t!]
\begin{center}
\includegraphics[width=8.cm, height=7.5cm]{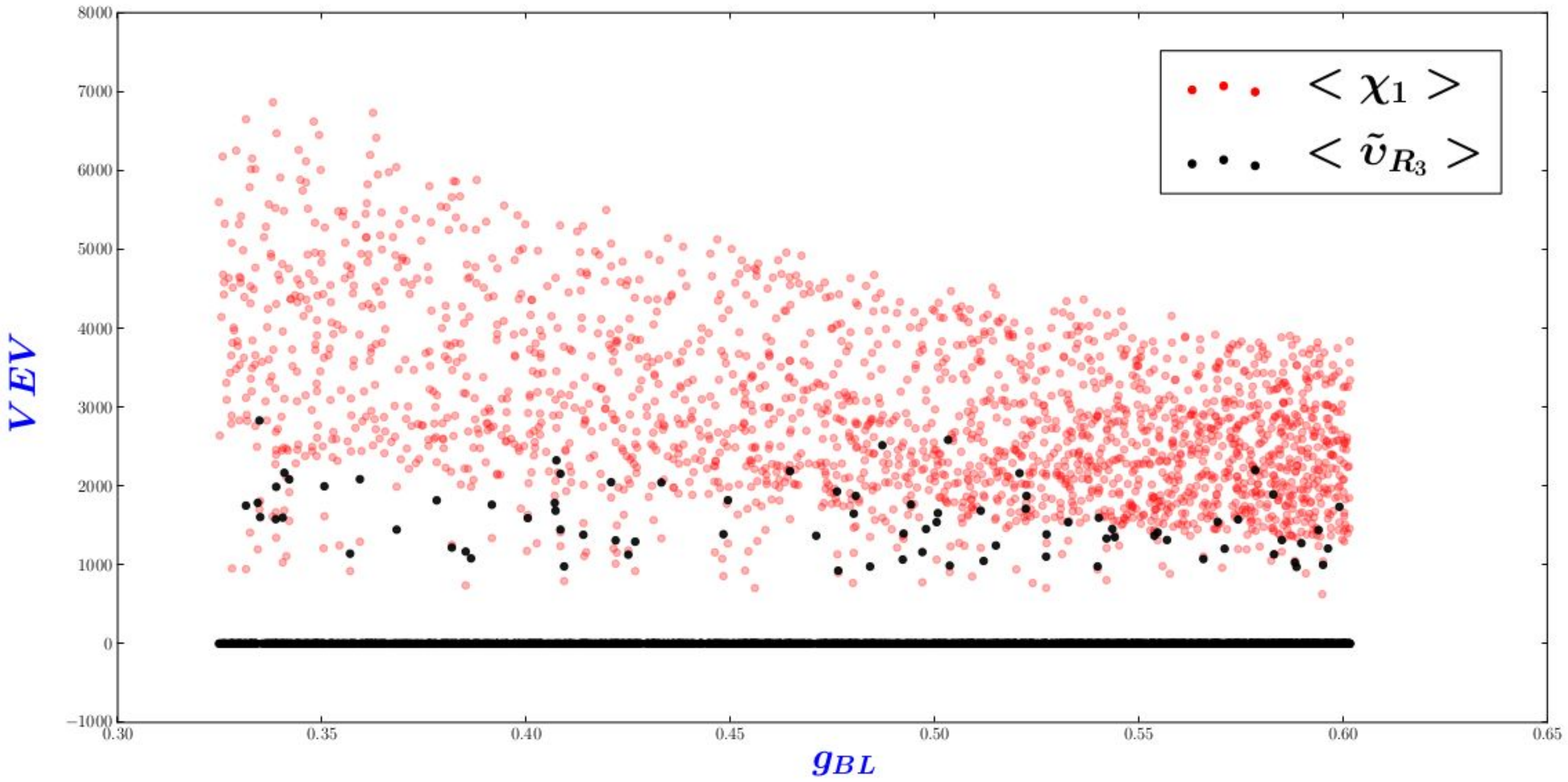}~~ \includegraphics[width=8.cm, height=7.5cm]{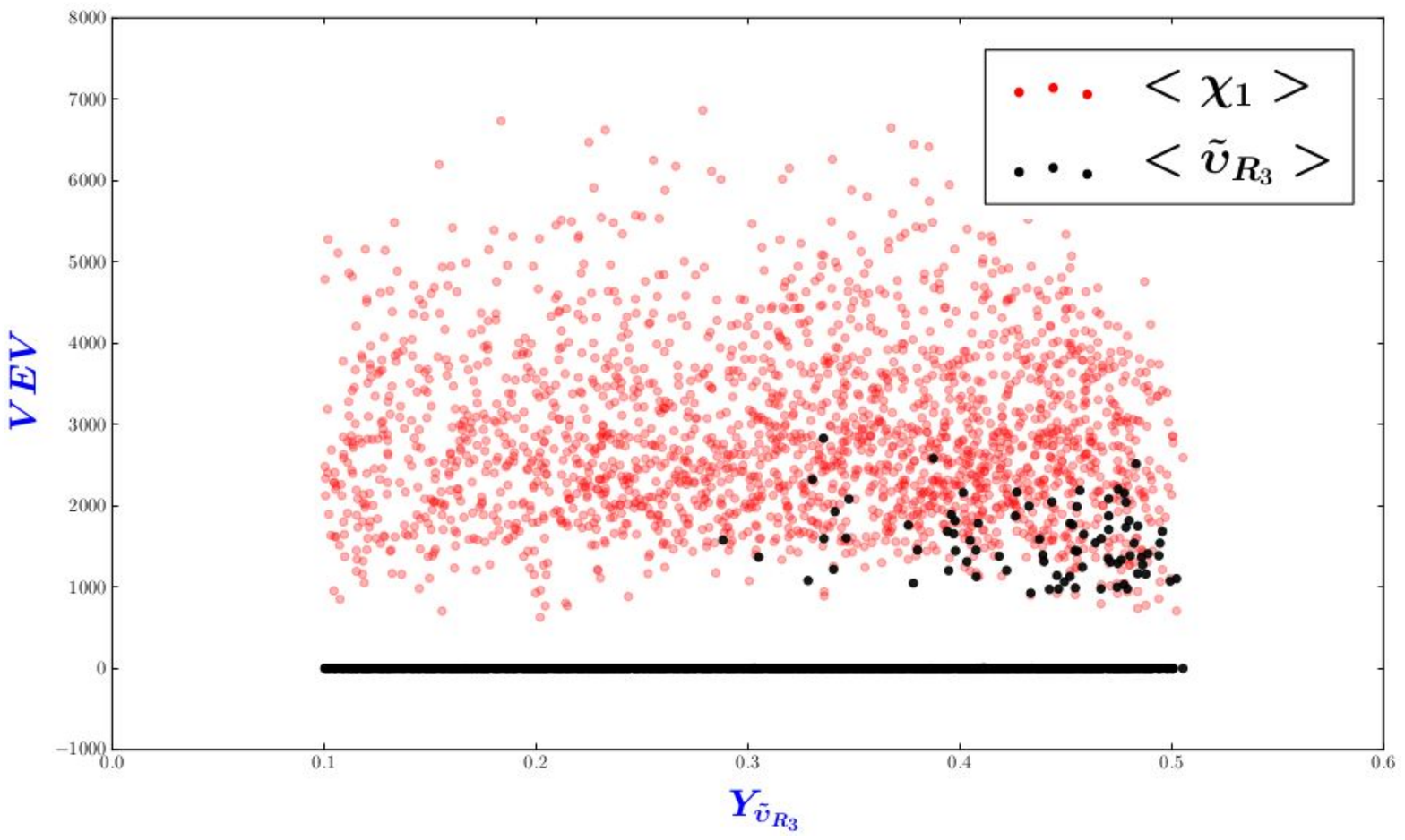}\\
\vskip -1cm
\includegraphics[width=8.cm, height=7.5cm]{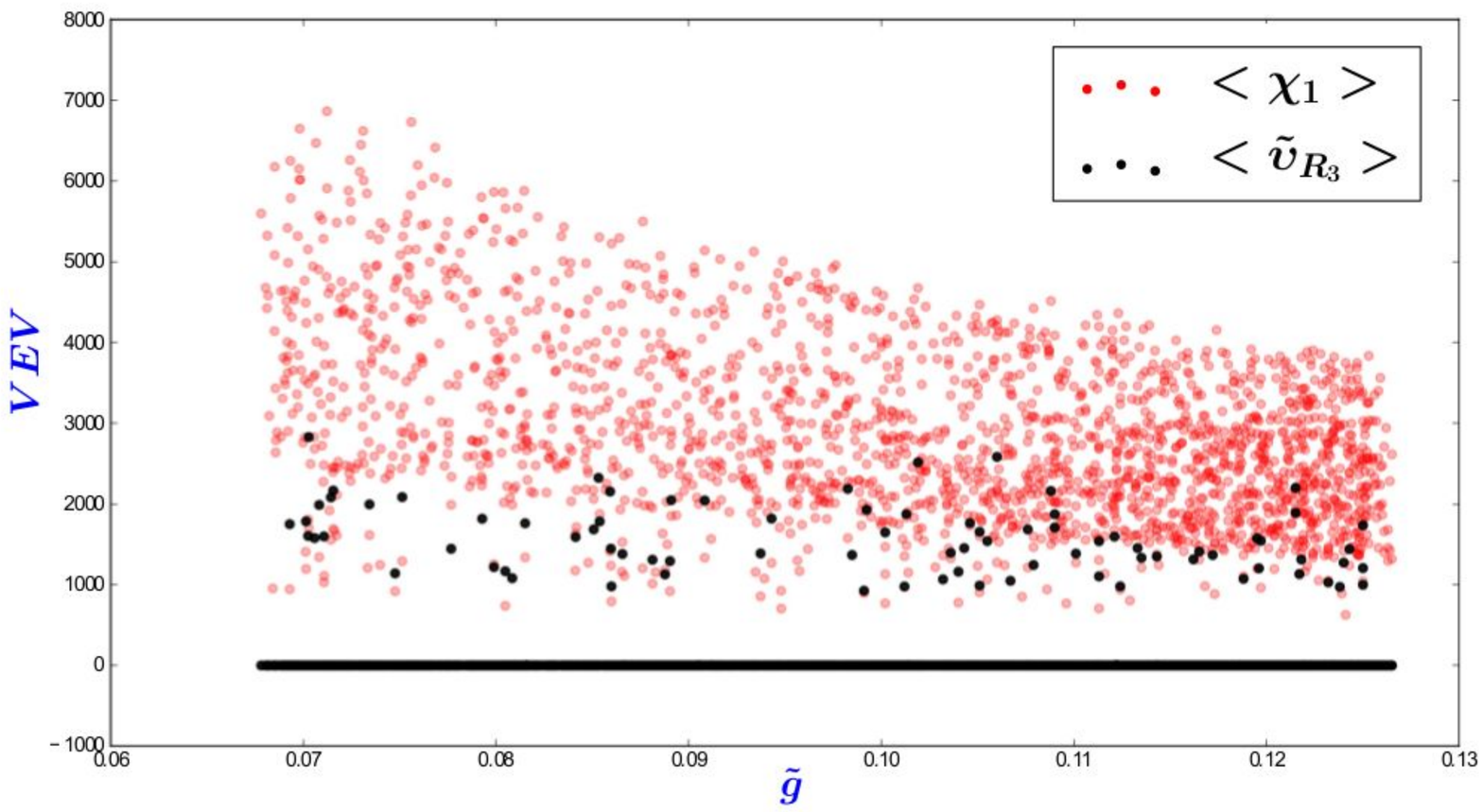}~~ \includegraphics[width=8.cm, height=7.5cm]{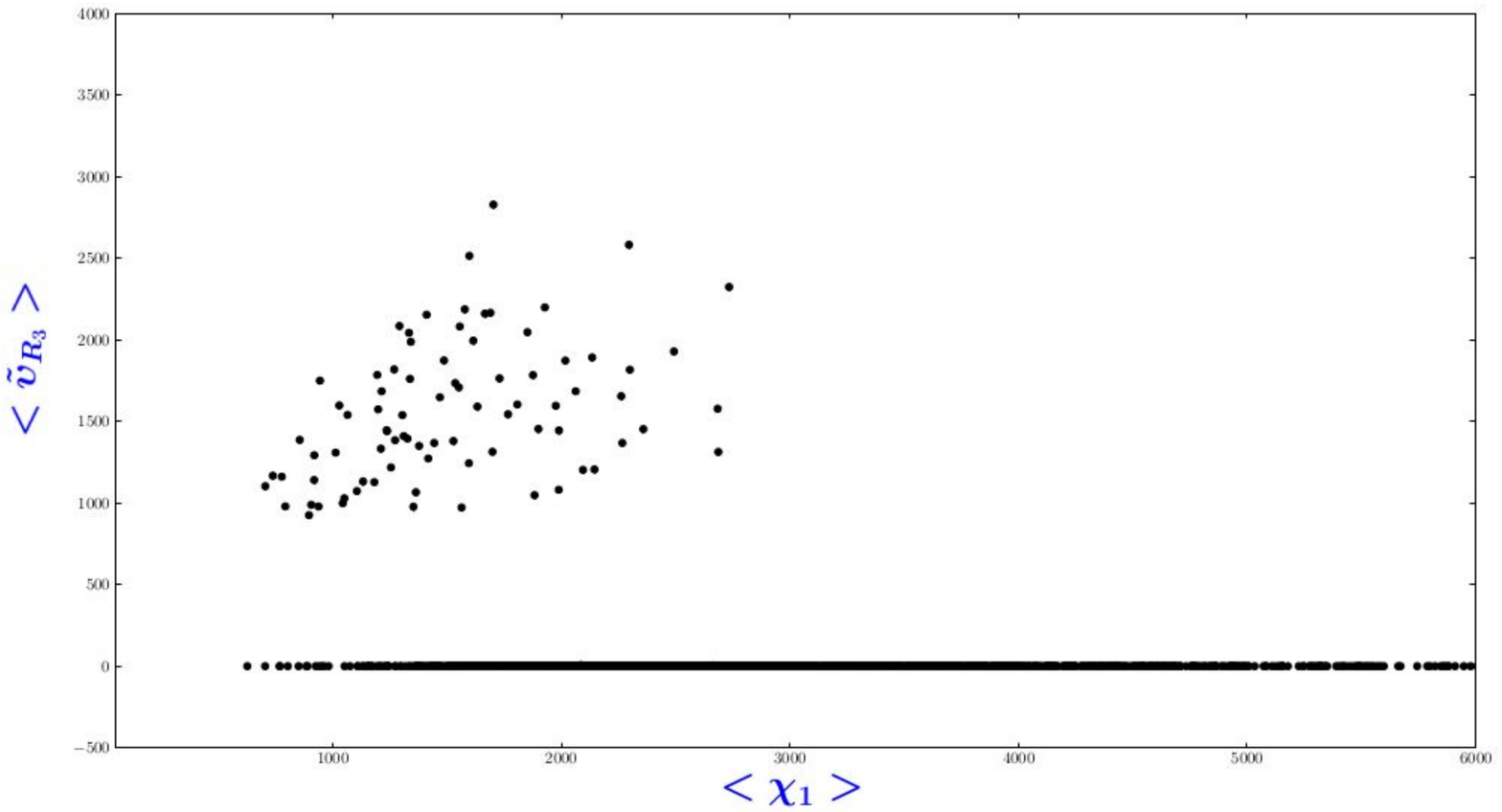}
\caption{The VEVs of singlet scalar $\chi_1$ and right-handed sneutrino $\tilde{\nu}_{R_3}$, in BLSSM-type I, as function of $g_{BL}$, $Y_{\nu_{R_3}}$and $\tilde{g}$. The last plot is for the correlation between these two VEVs.}
\label{BLSSM-vevs}
\end{center}
\end{figure*}

In Ref. \cite{CamargoMolina:2012hv}, it was shown that in BLSSM with type I seesaw, out of more than 2000 scanned points only about 100 points may lead to global non-zero vevs for neutrinos that may break the $R$-parity, while all other points preserve $R$-parity and break $B-L$ only. We confirmed these results by considering a wider range of parameter space. Our results are presented in Fig. \ref{BLSSM-vevs}, where the VEVs: $\langle \chi_1\rangle$  and $\langle \tilde{\nu}_{R_3} \rangle$ are given in terms of the relevant parameters $g_{BL}$, $Y_{\nu_{R_3}}$ and the gauge coupling mixing, $\tilde{g}$, between $U(1)_Y$ and $U(1)_{B-L}$. 
We also display the correlation between  $\langle \chi_1\rangle$  and $\langle \tilde{\nu}_{R_3} \rangle$. As can be seen from these plot, although most of the considered points lead to a non-vanishing $\langle \chi_1\rangle$  and zero $\langle \tilde{\nu}_{R_3} \rangle$, {\it i.e.}, $B-L$ is spontaneously broken while $R$-parity remains exact, there is a non-negligible number of points which induce non-vanishing $\langle \tilde{\nu}_{R_3} \rangle$, hence $R$-parity is broken along with the $B-L$. It is noticeable that the possibility of obtaining non-vanishing  $\langle \tilde{\nu}_{R_3} \rangle$ is increased with large values of $Y_{\tilde{\nu}_{R_3}}$ as found in previous section by the RGE evolution. In addition, the last plot in Fig.\ref{BLSSM-vevs} clearly shows that for most of the parameter space one gets  $\langle \chi_1\rangle = {\cal O}(1)$ TeV and $\langle \tilde{\nu}_{R_3} \rangle \simeq 0$. Nevertheless, it is quite plausible to have $\langle \tilde{\nu}_{R_3} \rangle \neq 0$ and  larger than $\langle \chi_1\rangle$. Those points will be the benchmarks of $R$-parity violation scenario studied in Ref. \cite{FileviezPerez:2010ek}.

\begin{figure*}[t!]
\begin{center}
\includegraphics[width=8.cm, height=6.5cm]{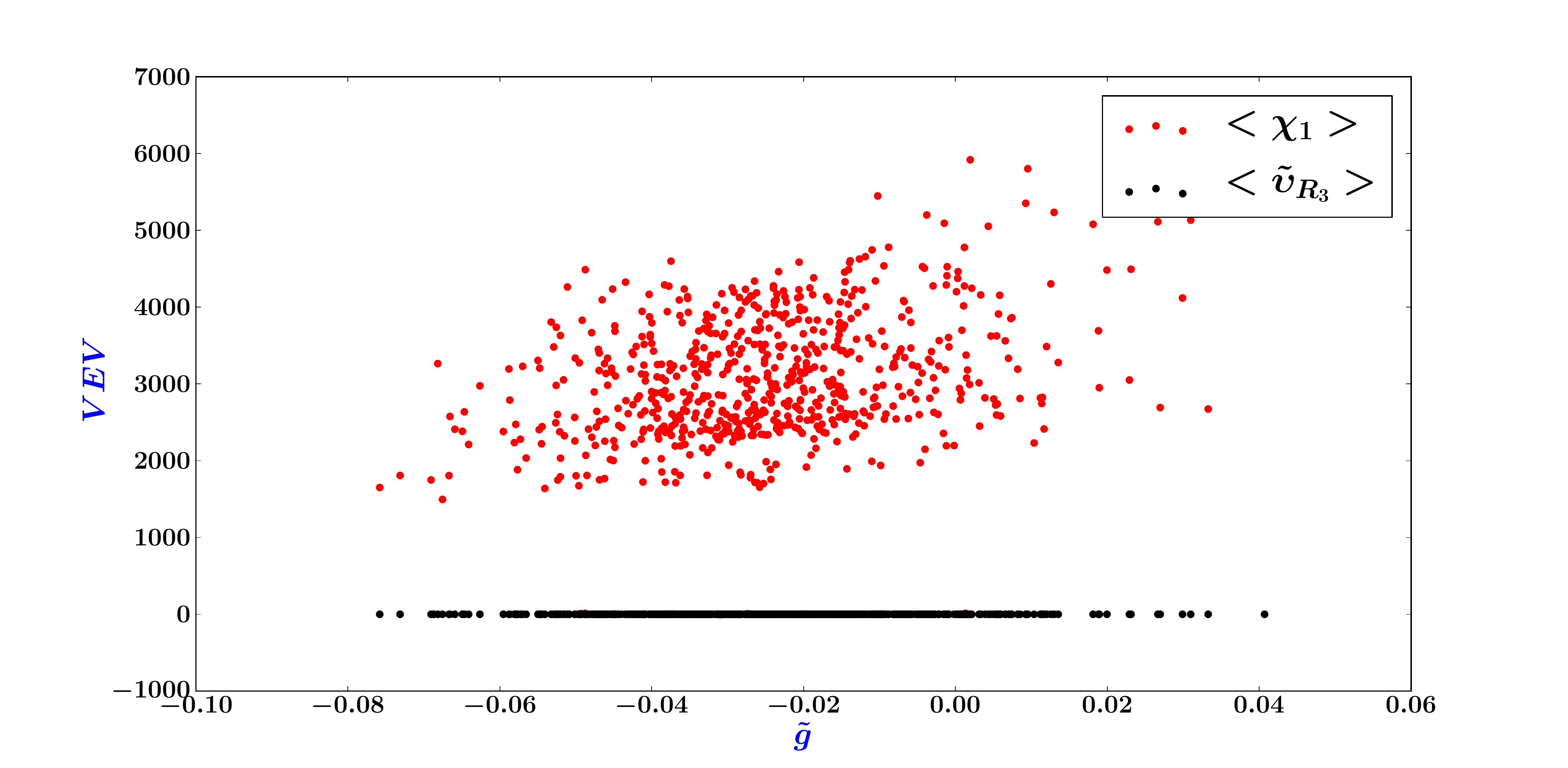}~~ \includegraphics[width=8.cm, height=6.5cm]{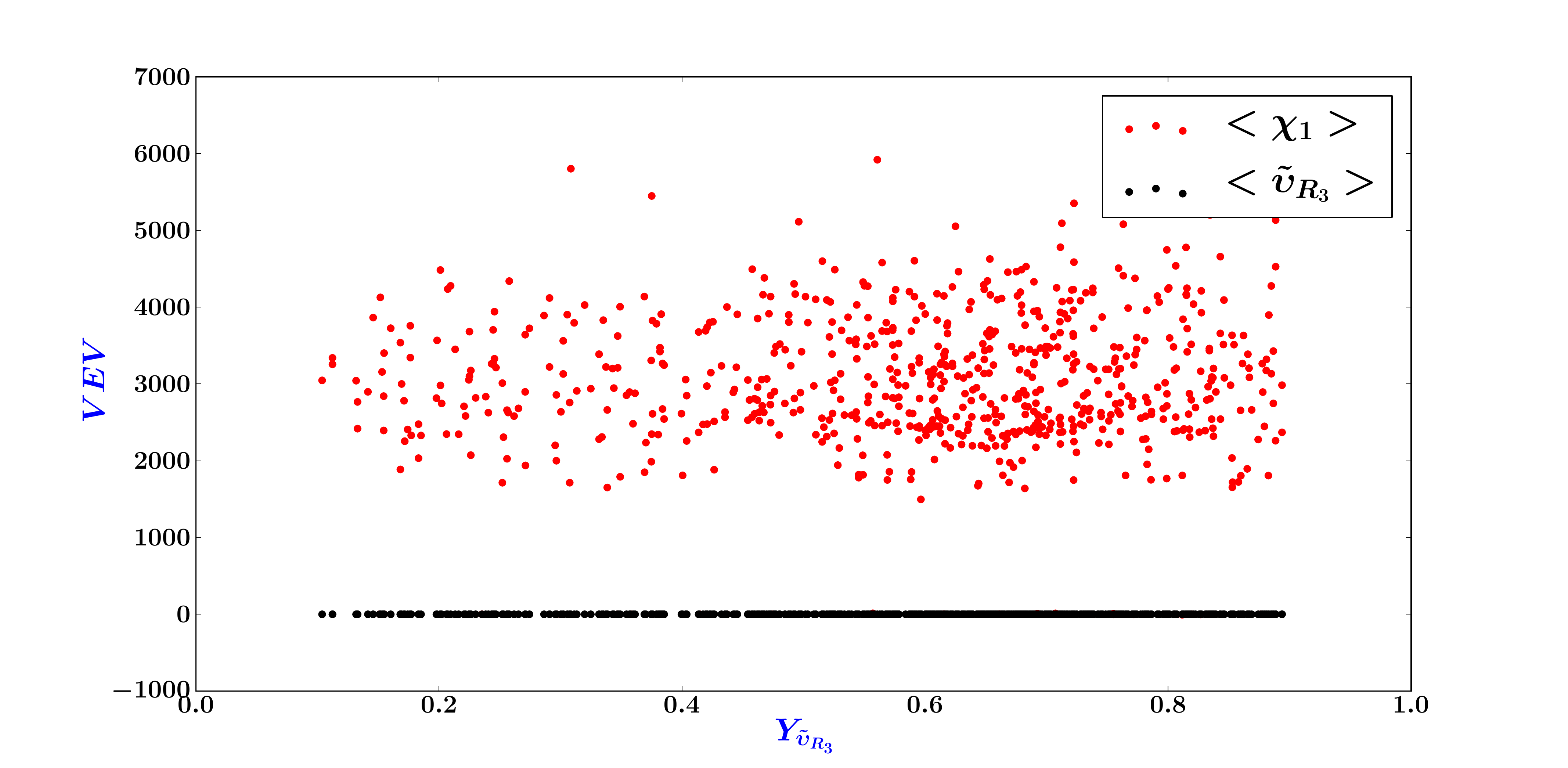}\\
\includegraphics[width=8.cm, height=6.5cm]{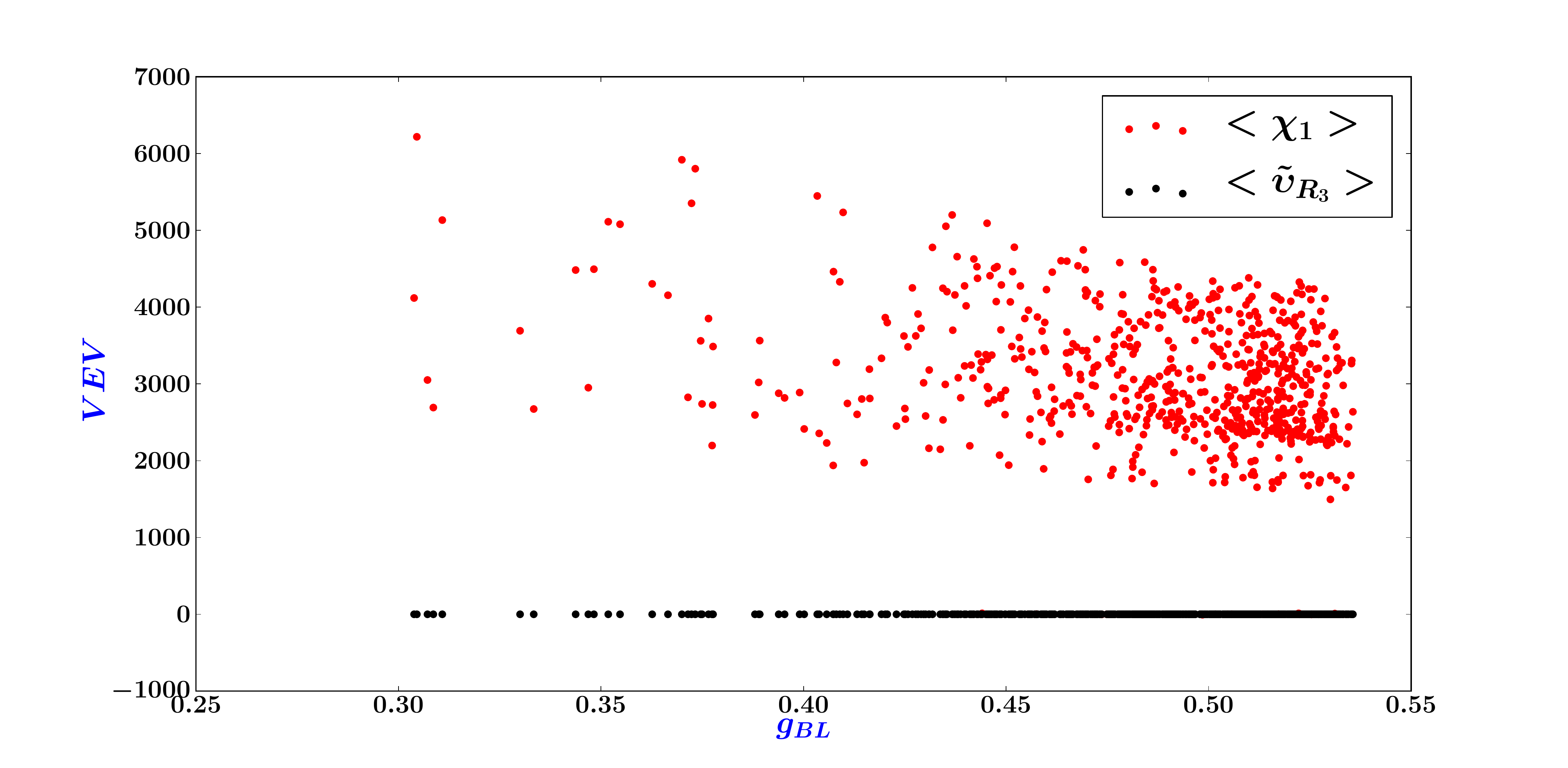}~~ \includegraphics[width=8cm, height=7.cm]{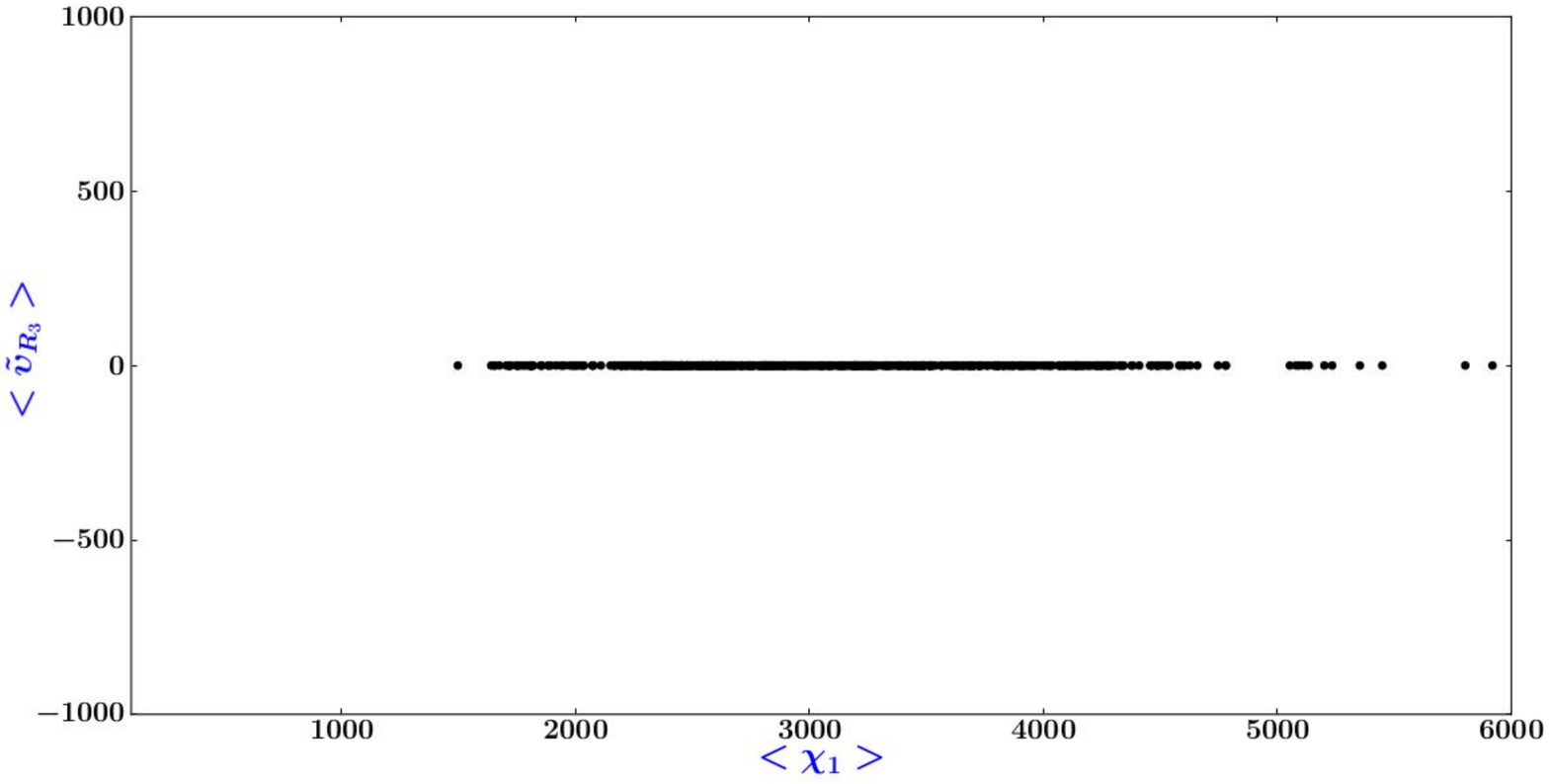}\\
\caption{The VEVs of singlet scalar $\chi_1$ and right-handed sneutrino $\tilde{\nu}_{R_3}$, in BLSSM-IS, as function of $g_{BL}$, $Y_{\nu_{R_3}}$and $\tilde{g}$. The last plot is for the correlation between these two VEVs.}
\label{BLSSMIS-vevs}
\end{center}
\end{figure*}

We performed a similar analysis for the BLSSM-IS.  We scanned over a large region of parameter space and checked the VEVs of the scalar fields $\chi_1$ and $\tilde{\nu}_{R_3}$. Fig. \ref{BLSSMIS-vevs} shows the resulting VEVS as functions of $\tilde{g}$, $Y_{\nu_{R_3}}$ and $g_{BL}$. Also the correlation between $\langle \chi_1\rangle$ and $\langle \tilde{\nu}_{R_3} \rangle$ is given in the last plot in this figure. It is now clear that unlike the case of BLSSM with type I seesaw, in BLSSM-IS there is no chance of getting non-zero VEV for the right-handed sneutrinos and one always finds $\langle \chi_1 \rangle \neq 0$ with $\langle \tilde{\nu}_{R_3}\rangle =0$. This conclusion is independent of the values of $g_{BL}$, $\tilde{g}$ or $Y_{\tilde{\nu}_{R_3}}$ as shown in Fig. \ref{BLSSMIS-vevs}. Moreover, the correlation between the VEVs in the last plot confirms that $\langle \chi_1 \rangle$ can be order TeV while $\langle \tilde{\nu}_{R_3}\rangle$ vanishes identically. Therefore, we can conclude that in the BLSSM-IS the $B-L$ symmetry can be radiatively broken while the $R$-parity remains as an exact symmetry.

\section{Conclusion}
\label{sec:summary}

We have analysed the radiative symmetry breaking of $B-L$ within the framework of BLSSM-IS. We considered the RGEs to show that for a wide range of parameters the squared mass of the Higgs singlet can be negative at TeV scale while the squared mass of the right-handed sneutrino remains positive.
Therefore, the $B-L$ symmetry is spontaneously broken by the VEV of this singlet and $R$-parity remains exact. We also investigated the vacuum stability of the BLSSM-IS, using the program of Vevacious. We showed that for a wide region of parameter space the singlet scalar $\chi_1$ gets a non-vanishing VEV $\sim {\cal O}(1)$ TeV and $\langle \tilde{\nu}_{R_3} \rangle =0$ so that $B-L$ is spontaneously broken and $R$-parity is conserved. This conclusion is different from the results obtained in BLSSM-type I, where $R$-parity can be spontaneously broken for a non-negligible number of points in parameter space.  
    
\vspace*{0.45cm}

\noindent
{\bf Acknowledgements}

\noindent
I would like to thank A. Hammad for his valuable help in using the Vevacious code and C. Un for fruitful discussions. 
This work was partially supported by the STDF project 13858 and the grant H2020-MSCA-RISE-2014 n. 645722 (NonMinimalHiggs).


\end{document}